\documentclass[11pt]{JHEP3}
\usepackage{amsmath}
\usepackage{amssymb,amsfonts}
\usepackage{layout}

\def\ni{\noindent}
\def\ga{\gamma}
\def\Ga{\Gamma}

\def\nn{\nonumber}

\def\a{\alpha}

\def\d{{\partial}}
\def\db{\bar{\partial}}
\def\zb{\bar{z}}
\def\wb{\bar{w}}

\def\uk{\frac{1}{\sqrt{k}}}
\def\utk{\frac{2}{\sqrt{k}}}

\def\b#1{{\bf #1}}
\def\bb#1{\bar{{\bf #1}}}
\def\ss#1{{\sf #1}}
\def\bp{{\bf p}}

\def\bg{{\bf g}}
\def\bv{{\bf v}}
\def\ba{\boldsymbol{\alpha}}

\def\lbldef#1#2{\expandafter\gdef\csname #1\endcsname {#2}}

\def\href#1#2{#2}

\newcommand{\beq}{\begin{equation}}
\newcommand{\eeq}{\end{equation}}
\newcommand{\ber}{\begin{eqnarray}}
\newcommand{\eer}{\end{eqnarray}}

\def\be{\begin{eqnarray}}
\def\ee{\end{eqnarray}}

\newcommand{\br}{\bar{r}}

\title{FZZ Scattering}
\preprint{hep-th/0606067 \\YITP-SB-06-20}

\author{Ari Pakman\footnote{Email: ari.pakman@stonybrook.edu} \\
\it C.N.~Yang Institute for Theoretical Physics,\\
\it Stony Brook University, \\
\it Stony Brook, NY 11794-3840, USA}

\abstract{We study the duality between the two dimensional black hole and the sine-Liouville
conformal field theories via exact
operator quantization of a classical scattering problem.
The ideas are first illustrated in Liouville theory, which is dual to itself
under the interchange of the Liouville parameter  $b$ by  $1/b$.
In both cases, a classical scattering problem does not determine
uniquely the quantum reflection coefficient. The latter is only fixed
by assuming that the dual scattering problem has the same reflection coefficient.
We also discuss the relation of this approach with the  method that exploits the
parafermionic symmetry of the model to compute the reflection coefficient.}


\begin{document}


\ni

\section{Introduction}
In an unpublished work \cite{FZZ}, Fateev, Zamolodchikov and Zamolodchikov discovered   that
 the \linebreak $SL(2,R)/U(1)$ WZW model
describing the two-dimensional conformal field theory whose target space is  the cigar,
or Euclidean 2D black hole \cite{Elitzur:1991cb,Mandal:1991tz,Witten:1991yr}
\be
\frac{ds^2}{k} = dr^2 + \tanh^2 r d\theta^2 \,, \qquad r>0, \qquad \theta \sim \theta + 2\pi
\label{cigarmetric}
\ee
where $k$ is the level of $SL(2,R)$,
has a dual description in terms of  a sine-Liouville theory.
This is a free theory with linear dilaton perturbed  by a sine-Liouville potential which
carries a unit of winding,
\be
{\cal L} = (\d \phi)^2 + (\d X)^2 + \mu \cos(\sqrt{k}\tilde{X})e^{-\sqrt{k'}\phi} + {\cal R}\,  k'^{-1/2}\phi \,,
\label{lagrangian}
\ee
where  $k'=k-2$ and $\tilde{X}= X_L-X_R$.

This so-called FZZ duality, has been investigated and exploited  in several
works \cite{Baseilhac:1998eq,Kazakov:2000pm, Fukuda:2001jd, Giveon:2001up,
Giribet:2004zd,Hikida:2004mp, Bergman:2005qf, Mukherjee:2006zv}.
It has been generalized to the $N=2$ supersymmetric case \cite{Giveon:1999px}, where
it follows from mirror symmetry~\cite{Hori:2001ax} (see also \cite{Tong:2003ik}).
For worldsheets with boundary  there exists a boundary version of the FZZ duality, both when
the bulk theory has the cigar/sine-Liouville perturbations (or their $N=2$ counterparts)
turned on~\cite{Israel:2004jt}
or when the bulk theory is flat. In the latter case, the D1  brane of the cigar~\cite{Ribault:2003ss} becomes
the hairpin brane~\cite{Lukyanov:2003nj}, which has a sine-Liouville dual description studied in~\cite{Kutasov:2005rr,Lukyanov:2005bf}.

In this paper we explore the  FZZ duality by studying
the exact operator quantization of a classical scattering problem.
The same ideas and techniques are relevant for the simpler case of  Liouville
theory, which we discuss first as a warm up.
In both cases we show that the scattering coefficient of a free field
bouncing off the Liouville wall or the tip of the cigar cannot be determined without further input,
which comes from assuming that there is a second scattering process with the same
scattering coefficient. In the case of Liouville, discussed in Section 2,
the second theory is Liouville
itself with the Liouville coefficient $b$ replaced by $b^{-1}$. In
the case of the cigar, discussed in Section 3, we find a field which represents a scattering process of sine-Liouville type
 and which  yields the correct reflection coefficient,   computed
previously using other techniques. We also comment on the relation between the results of this work and
the techniques considered in \cite{Mukherjee:2006zv}, which are based on exploiting the
parafermionic symmetry of  the model.

\section{Duality in Liouville scattering  }
It is well known that the Liouville classical equation can be solved in terms of a free field.
Therefore it is natural to try to quantize Liouville theory by quantizing the mapping to this free field via operator quantization.
This program has been carried out successfully in \cite{Teschner:2003en} were the DOZZ formula \cite{Dorn:1994xn,Zamolodchikov:1995aa}
for the Liouville
three point function was reobtained (see also \cite{Jorjadze:2003cg}).

In this section
we will use similar techniques to compute the reflection coefficient of an asymptotically free field
bouncing off  the Liouville wall. The relevant ideas were
 succinctly exposed in \cite{Teschner:2001rv} (see Section 14 there), and we will flesh them out here,
 stressing the role of the $b \leftrightarrow 1/b$ duality.

\subsection{Classical scattering}
We work first in the cylinder $(\sigma,t) \sim (\sigma + 2 \pi, t)$.
The Liouville equation of motion
\be
\d_+ \d_-  \varphi(\sigma,t) = 2 \pi \mu_c b  e^{b \varphi(\sigma,t)}\,,
\ee
where $x^{\pm}=t \pm  \sigma $, can be solved in terms of two arbitrary functions $\mathrm{B}=\mathrm{B}(x^+),
\bar{\mathrm{B}} = \bar{\mathrm{B}}(x^-)$,
\be
 e^{- \frac{b {\varphi }}{2}} = \sqrt{\pi \mu_c} b \, 
 \frac{ 1 - \mathrm{B} \bar{\mathrm{B}} }{\sqrt{\d_+ \mathrm{B} \d_- \bar{\mathrm{B}}}} \,.
\label{lsolution}
\ee
It is convenient to express $\mathrm{B}, \bar{\mathrm{B}}$
through a free field $\phi(\sigma,t)= \phi(x^+) + \bar{\phi}(x^-)$ as
\be
\d_{+} \mathrm{B} &=& \sqrt{\pi \mu_c} b e^{b \phi(x^+)} \,,
\label{a}\\
\d_{-} \bar{\mathrm{B}} &=& \sqrt{\pi \mu_c} b e^{b \bar{\phi}(x^-)} \,.
\label{abar}
\ee
The fields $\phi, \bar{\phi}$ have an expansion,
\be
\phi(x^+) &=& \frac{x}{2} +\frac{p}{2}x^+
+ \textrm{oscilators}
\\
\bar{\phi}(x^-) &=& \frac{x}{2} +\frac{p}{2}x^-
+\textrm{oscilators}
\ee
Requiring $\mathrm{B}$ to have the same monodromy as $\d_+\mathrm{B}(x^+ )$, $\d_+\mathrm{B}(x^+ + 2\pi) = e^{\pi p b}\d_+\mathrm{B}(x^+)$,
and similarly for $\bar{\mathrm{B}}$,
fixes the solutions to (\ref{a})-(\ref{abar}) as
\be
\mathrm{B}(x^+) &=& \frac{\sqrt{\pi \mu_c} b}{(e^{\pi p b} -1)} \int_0^{2\pi} \!\! d \sigma' e^{b \phi(x^+ + \sigma')} \,,
\\
\bar{\mathrm{B}}(x^-) &=& \frac{\sqrt{\pi \mu_c} b}{(e^{\pi p b} -1)} \int_0^{2\pi} \!\! d \sigma'' e^{b \phi(x^- + \sigma'')} \,.
\ee
The solution (\ref{lsolution}) is  invariant
under $\mathrm{B},\bar{\mathrm{B}} \rightarrow 1/\mathrm{B}, 1/\mathrm{\bar{B}}$.
This transformation corresponds to mapping  the
free field $\phi(\sigma,t)$ into another free field $\xi(\sigma,t) = \xi(x^+) + \xi(x^-)$ with momentum $-p$,
given by
\be
e^{-\frac{b}{2} (\xi(x^+) + \xi(x^-)) } = -\mathrm{B}(x^+)\mathrm{\bar{B}}(x^-) e^{-\frac{b}{2} (\phi(x^+)+  \phi(x^-))}\,.
\ee
This mapping encodes the physical meaning of the solution (\ref{lsolution})
as a  scattering of the free field~$\phi(\sigma,t)$ off the Liouville wall \cite{Dzhordzhadze:1979qj}.
To see this, note first that as a function of $\phi,\bar{\phi}$, eq.(\ref{lsolution})  is
\be
e^{-\frac{b\varphi(\sigma,t) }{2}} = e^{-\frac{b}{2}(\phi(x^+)+ \phi(x^-))} \left[ 1 -  \frac{\pi \mu_c b^2}{(e^{\pi p b}-1)^2}
\int_0^{2\pi} \!\! d \sigma' e^{b \phi(x^+ + \sigma')}  \int_0^{2\pi} \!\! d \sigma'' e^{b \phi(x^- + \sigma'')}  \right] \,.
\label{map}
\ee
Suppose that~$p>0$. Then at the infinite past and future we have
\be
\lim_{t \rightarrow - \infty} e^{-\frac{b \varphi(\sigma,t) }{2}} &=& e^{-\frac{b}{2}(\phi(x^+)+ \bar{\phi}(x^-))}\,,  \\
\lim_{t \rightarrow + \infty} e^{-\frac{b \varphi(\sigma,t) }{2}} &=& e^{-\frac{b}{2} (\xi(x^+) + \xi(x^-)) } \,.
\ee
Therefore at both ends ${t \rightarrow \pm \infty}$ the Liouville field $\varphi$ is a free field. The Liouville wall
maps  the incoming ($p>0$) free field $\phi(\sigma,t)$ into the outgoing
($p<0$) free field $\xi(\sigma,t)$.
The case~$p<0$ is obtained by time reversal.

\subsection{The reflection coefficient and the Liouville duality}
A remarkable fact is that the transformation from the field $\phi(\sigma,t)$
to the field $\varphi(\sigma,t)$ is canonical~\cite{Teschner:2001rv}. This suggests to quantize Liouville
theory by quantizing the free field $\phi(\sigma,t)$ and defining a quantum
version of eq.(\ref{map}). We will perform the quantization in the complex
Euclidean plane. Let us Wick rotate the time variable into
$\tau= it$, and take  $z=e^{\tau + i \sigma}, \zb=e^{\tau - i \sigma}$ as two independent variables.

The fields $\phi, \bar{\phi}$ have a mode expansion
\be
\phi(z) &=& \b{x} -i\frac{\b{p}}{2}\log z
+ \frac{i}{\sqrt{2}} \sum_{n \neq 0} \frac{\alpha_n}{n}\frac{1}{z^n} \,,
\\
\bar{\phi}(\bar{z}) &=& \b{x}-i\frac{\b{p}}{2}\log \bar{z}
+ \frac{i}{\sqrt{2}} \sum_{n \neq 0} \frac{\bar{\alpha}_n}{n} \frac{1}{\bar{z}^n} \,,
\ee
and the modes
satisfy the commutation relations
\be
\left[ \b{x},\b{p} \right] & = & i \\
\left[ \a_n , \a_m \right] &=&  \left[ \bar{\a}_n , \bar{\a}_m \right] = n  \delta_{n+m,0} \,.
\ee
Normal ordering $::$ is defined by putting the annihilation operators $\alpha_{n>0}$ and $\b{p}$
 to the right, and the creation operators $\alpha_{n<0}$
and $\b{x}$  to the left.
With this prescription the short distance singularity is
\be
\phi(z) \phi(w) &=& :\phi(z) \phi(w): - \frac12 \log(z-w) \,,
\ee
and similarly for $\bar{\phi}(\zb)$.
The stress tensor of the quantum theory is
\be
T(z) = -(\d \phi(z))^2 +Q \d^2 \phi(z)   \,,
\ee
with
\be
Q = b + b^{-1}
\ee
and central charge
\be
c= 1 + 6Q^2 \,,
\ee
and there is a similar anti-holomorphic copy.
A chiral vertex operator of the form
\be
 :e^{2 \a \phi(z)} :  \,
 = e^{2 \a \b{x}} e^{-2\a\frac{i \b{p}}{2} \log z }
e^{-\frac{i 2\a}{\sqrt{2}}{\sum}_{n > 0} \frac{\alpha_{-n}}{n}{z^n}}
e^{\frac{i 2\a}{\sqrt{2}} \sum_{n > 0} \frac{\alpha_n}{n}{z^{-n}}} \,,
\ee
has holomorphic conformal
dimension
\be
\Delta = \a(Q-\a) \,,
\label{confdim}
\ee
and carries $\b{p}$ momentum
\be
p = -2i\a \,.
\label{palpha}
\ee
For normalizable states with $\a = \frac{Q}{2} + i\mathbb{R}$,
the last expression is  consistent
with the anomalous hermiticity  of $\b{p}$,
\be
\bp{\dagger} = \bp + 2iQ\,,
\ee
which follows from the presence of the background charge.

In order to define the quantum version of  $e^{-\frac{b \varphi}{2}}$,
let us define
the ``screening charges''
\be
\b{S}(z) &=& \int_{z}^{z e^{2\pi i}} \!\!\!\!\!\!\!\!\!\!\! dw \,\, e^{2b \phi(w)} \,, \\
\bar{\b{S}}(\zb) &=& \int_{\zb}^{\zb e^{2\pi i}} \!\!\!\!\!\!\!\!\!\!\! d \wb \,\, e^{2b \bar{\phi}(\wb)} \,.
\ee
Since the integrand has conformal dimension one, these are conformal primaries with   dimension zero, as follows from
\be
T(y)\b{S}(z) &\sim&
\int_z^{ze^{2\pi i}} \!\!\!\!\!\!\! dw \,
\frac{\d}{\d w}
\left[ \frac{e^{2b \phi(w)} }{y-w}\right] \,,
\label{dimscreen} \\
&\sim&
\frac{(e^{2 b \pi (\b{p}+2ib)} -1)}{y-z}
e^{2b \phi(z)} \,, \nn \\
&\sim& \frac{1}{y-z}
\frac{\d \b{S} (z)}{\d z} \,.
\nn
\ee
The quantum version of $e^{-\frac{b \varphi}{2}}$ in (\ref{map}) can now be taken  as
\be
\b{V}_b(z,\zb) = e^{-b \phi(z)} e^{b \b{x}} e^{-b \bar{\phi}(\zb)}
-  \frac{\mu_c \pi b^2}{(e^{-4 \pi i b(\ba-b)}-1)}
\b{S}(z) e^{-b \phi(z)} e^{-b \b{x}} e^{-b \bar{\phi}(\zb)}
\bar{\b{S}}(\zb)
\frac{1}{(e^{-4 \pi i b\ba}-1)} \,,
\ee
where we have defined   $\ba = i \frac{ \b{p}}{2}$ (see (\ref{palpha})). The
factors $e^{\pm b \b{x}}$ are inserted in order to eliminate the duplication of the
zero modes coming from $\phi$ and $\bar{\phi}$. Their position is dictated by the
preservation of the normal order in the exponentials.
In the factor  $(e^{-4 \pi i b(\ba-b)}-1)^{-1}$, the shift in $\ba$ is  because
we have written it to the left of $\b{S}(z) e^{-b \phi(z)}$.

An important property of the operator $\b{V}_b$, is that, when defined in the Minkowskian
cylinder~$(t,\sigma)$, it satisfies the locality condition \cite{Teschner:2001rv}
\be
[\b{V}_b(t,\sigma), \b{V}_b(t, \sigma')]=0 \,.
\label{loca1}
\ee
The product of screening charges and exponentials in  $\b{V}_b$ is
\be
 \b{S}(z) e^{-b \phi(z)} =
\int_z^{ze^{2\pi i}} \!\!\!\!\!\!\! dw \, (w-z)^{b^2} :e^{2b \phi(w) - b\phi(z)}: \,,
\label{prod1}
\\
e^{-b \bar{\phi}(\zb)}  \bar{\b{S}}(\zb)  =
\int_{\zb}^{\zb e^{2\pi i}} \!\!\!\!\!\!\! d \wb \, (\zb-\wb)^{b^2} :e^{2b \bar{\phi}(\wb) - b\bar{\phi}(\zb)}: \,.
\label{prod2}
\ee
In the interacting theory, the  primary fields can still be labeled by $\a$, with conformal dimension $\a(Q-\a)$.
This expression   is symmetric under $\a \rightarrow Q-\a$.
For delta-normalizable states with $\a=Q/2 + i P$, this corresponds to $P \rightarrow -P$.
Therefore an operator $\ss{S}$ which maps a state with $\a$ into one with $Q-\a$ can be considered
as  the quantum version of the mapping between asymptotic free fields provided by the classical solution.
We do not need here the explicit form of the Liouville primaries for arbitrary $\a$ (see \cite{Teschner:2001rv,Teschner:2003en}).
It is enough for us that their action on the vacuum creates a state $|\a\rangle$, such that
\be
\ss{S}|\a \rangle = |Q-\a\rangle \,.
\ee
The ambiguity of the parametrization of the classical solution using $\phi$ or $\xi$,
should manifest itself in the invariance
\be
\b{V}_{b}(z,\zb) = \ss{S}^{-1} \b{V}_{b}(z,\zb) \ss{S}\,.
\label{invariance}
\ee
The operator $\ss{S}$ should be a  product $\ss{S}=\ss{P}\ss{R}$, where the operator $\ss{P}$
acts on the zero modes to change the eigenvalue of $\ba$, and $\ss{R}$
acts as
\be
\ss{R}|\a \rangle = R(\a) |\a \rangle\,,
\ee
where $R(\a)$
is the reflection coefficient that we want to compute. Note that from $\ss{S}^2=1$ it follows that
\be
R(\a) R(Q-\a) =1\,.
\label{rlinv}
\ee
An arbitrary matrix element of $\b{V}_{b}(z,\zb)$ at $z=\zb=1$ is,
taking proper care of the zero modes,
\be
\langle \a'| \b{V}_b(1,1) | \a \rangle = \delta(\a' - \a + b/2) + \delta(\a' - \a - b/2) D_b(\a)\,.
\label{matrixv}
\ee
Using (\ref{prod1})-(\ref{prod2}), the function $D_b(\a)$ is given by
\be
D_b(\a) = - \mu_c \pi b^2 \frac{1}{(e^{-4 \pi i b(\a-b/2)}-1)}
I(\a) \bar{I}(\a) \frac{1}{(e^{-4 \pi i b \a}-1)}\,,
\ee
where
\be
I(\a) &=& \oint dw (w-1)^{b^2} w^{-2b\a} \,, \\
\bar{I}(\a) &=& \oint d \wb (1-\wb)^{b^2} \wb^{-2b\a} \\
&=& e^{-i\pi b^2} I(\a) \,,
\ee
and both integrals are taken counterclockwise in the unit circle.
Since the integrand of $I(\a)$ is not analytic at $w=0,1$, the contour can be deformed
keeping the point $w=1$ fixed and without crossing the point $w=0$. This leads to
\be
I(\a)&=& e^{\pi i b^2}(e^{-4\pi i b}-1) \int^{1}_0 \!\!dw \, (1-w)^{b^2} w^{-2b\a} \,, \nn \\
&=& e^{\pi i b^2}(e^{-4\pi i b}-1)  \frac{\Gamma(1-2b \a ) \Gamma(b^2 +1)}{\Gamma( 2 -2b \a + b^2)},
\ee
where we used the integral representation (\ref{euler}) of the Euler beta function.
The final expression for $D_b(\a)$ is
\be
D_b(\a) = \mu_c b^2 \pi \Gamma^2(1+b^2) \ga(2 \a b - b^2 -1) \ga(1 - 2\a b)\,,
\ee
where $\ga(x)=\Ga(x)/\Ga(1-x)$. From the invariance (\ref{invariance}), it follows that equation (\ref{matrixv}) should be equal to
\be
\langle \a'|  \ss{S}^{-1} \b{V}_{b}(1,1) \ss{S} | \a \rangle &=&
R^{-1}(\a +b/2) R(\a) \delta(\a' - \a -b/2)\nn \\ &&
\qquad  + R^{-1}(\a -b/2)R(\a) D_b(Q-\a) \delta(\a' - \a +b/2) \,.
\label{matrixvs}
\ee
Comparing the coefficients of the delta functions in (\ref{matrixv}) and (\ref{matrixvs})
gives  two equations for $R(\a)$, but it is easy to see that
they are equivalent  using  (\ref{rlinv}).
The resulting equation for $R(\a)$ is
\be
R(\a + b/2) = R(\a) D_b^{-1}(\a) \,.
\label{liouvilleshift}
\ee
This is a difference equation that constraints the form of $R(\a)$ but does not fix it uniquely,
since any solution can be multiplied by an arbitrary periodic function of $\a$ with period $b/2$.

So we find that the classical Liouville scattering problem has no unique quantum version.
A similar  ambiguity is encountered in the bootstrap approach to quantum Liouville theory \cite{Teschner:1995yf,Pakman:2006hm}.
There, one imposes the symmetry $b \leftrightarrow 1/b$. In our case this leads to a second equation for $R(\a)$,
\be
R(\a + b^{-1}/2) = R(\a) \tilde{D}_{1/b}^{-1}(\a) \,,
\label{liouvilleshift2}
\ee
where
\be
\tilde{D}_{1/b}(\a) = \tilde{\mu}_c b^{-2} \pi \Gamma^2(1+b^{-2}) \ga(2 \a b^{-1} - b^{-2} -1) \ga(1 - 2\a b^{-1})\,.
\ee
Eqs.(\ref{rlinv}),  (\ref{liouvilleshift}) and (\ref{liouvilleshift2}), yield, for irrational $b^2$, a unique reflection coefficient,
\be
R(\a) = -\big ( \mu_c b^2 \Gamma^2(b^2)  \big)^{\frac{Q-2\a}{b}} \left[ (Q-2\a )^2 \ga(b(Q-2\a)) \ga(b^{-1}(Q-2\a) ) \right]^{-1} \,,
\ee
with $\tilde{\mu}_c b^{-2} \Gamma^2(b^{-2}) = \left(\mu_c b^2 \Gamma^2(b^2) \right)^{1/b^2}$.
This is the same reflection coefficient that follows from the DOZZ formula \cite{Dorn:1994xn,Zamolodchikov:1995aa}, with the
identification
\be
\mu_c \pi b^2  = \mu_{_{DOZZ}} \sin(\pi b^2)\,.
\ee

\vskip 0.3 cm
\ni
The important lesson that we learn is that in order to fix uniquely
the reflection coefficient of Liouville theory, we must assume that there are {\it two} classical
scattering problems (related by $b \leftrightarrow 1/b$), with  the {\it same} quantum reflection coefficient.

\section{FZZ scattering }
In this section we will address a similar scattering problem formulated in
the cigar background~(\ref{cigarmetric}).
The classical equations of the cigar non-linear sigma model
were solved in  terms of free fields
 in~\cite{Mueller:1996qg,Muller:1998bd,Muller:1999xf}.
This solution was canonically quantized in the interesting  work~\cite{Kruger:2004jm}.
We start reviewing some results of these works.
As we will see, the equations obtained from quantizing the cigar are not enough
to fix the reflection coefficient. We will show that assuming that a dual
scattering problem of sine-Liouville type
has the same reflection coefficient, fixes it  to its known value from other quantization schemes.

\subsection{Classical scattering in the cigar}
Let us consider the cigar metric (\ref{cigarmetric}) parameterized with Kruskal  coordinates
\be
u = \frac{\sinh r e^{i \theta }}{\sqrt{\lambda}}
\qquad
\bar{u} = \frac{\sinh r e^{-i \theta }}{\sqrt{\lambda}}  \,.
\ee
The parameter $\lambda$ will play a role
similar to $\mu_c$  in Liouville theory.
In these variables, the classical action of a string propagating in the
cigar background (\ref{cigarmetric}) is, in the complex plane,
\be
S[u,\bar{u}] = \frac{k}{4 \pi} \int dz d\bar{z}
\left[ \frac{\d u \db \bar{u} + \d \bar{u} \db u}{\lambda+ u \bar{u}}  \right]\, .
\ee
The equations of motion following from this action are the real and
imaginary parts of
\be
\d \db u = \frac{\bar{u}\d u \db u}{\lambda+ u \bar{u}}\,\,.
\label{EOM}
\ee
The exact solution to this equation  found in
\cite{Mueller:1996qg,Muller:1998bd,Muller:1999xf}
in terms of two free fields
$\phi(z,\zb)= \phi(z)+ \bar{\phi}(\zb)$ and $X(z,\zb)=X(z) + \bar{X}(\zb)$ reads
\be
u(z,\zb) = e^{\uk (\phi(z) + \bar{\phi}(\zb)+ i X(z) + i \bar{X}(\zb))}
\left[1 - \lambda A(z)\bar{A}(\zb)\right] \,,
\label{classu}
\ee
where $A(z)$ and $\bar{A}(\zb)$ are solutions to
\be
\d A(z) &=& \uk (\d \phi(z) + i \d X(z))e^{-\utk \phi(z)} \,, \\
\db \bar{A}(\zb) &=& \uk (\db \bar{\phi}(\zb) + i \db \bar{X}(\zb))e^{-\utk \bar{\phi}(\zb)} \,.
\ee
As in Liouville theory, this mapping to free fields is canonical \cite{Mueller:1996qg,Muller:1998bd,Muller:1999xf}.
Expanding the fields $\phi, \bar{\phi}$  in modes as
\be
\phi(z) &=& \frac{x_1}{2} -i\frac{p_1}{2}\log z
+ \textrm{oscilators}
\label{modephi}\\
\bar{\phi}(\bar{z}) &=& \frac{x_1}{2} -i\frac{p_1}{2}\log \bar{z}
+ \textrm{oscilators}
\label{modephibar}
\ee
we see that $\d A(z), \db \bar{A}(\zb)$
have monodromies
\be
\d A(e^{2\pi i} z ) &= & \d A(z) e^{- \frac{2 \pi p_1}{\sqrt{k}}}\,,  \\
\db \bar{A}(e^{2\pi i} \zb ) &=&  \db \bar{A}(\zb) e^{- \frac{2 \pi p_1}{\sqrt{k}}}\,.
\ee
Requiring these monodromies to be preserved by $A(z), \bar{A}(\zb)$,
fixes them uniquely as
\be
A(z) &=& \frac{1}{\sqrt{k}(e^{- \frac{2 \pi p_1}{\sqrt{k}}} -1)}
\int_z^{ze^{2\pi i}} \!\!\!\!\!\!\! dw \, (\d \phi(w) + i \d X(w))e^{-\utk \phi(w)} \,,
\\
\bar{A}(\zb) &=& \frac{1}{\sqrt{k}(e^{- \frac{2 \pi p_1}{\sqrt{k}}} -1)}
\int_{\zb}^{\zb e^{2\pi i}} \!\!\!\!\!\!\! d \bar{w} \,
(\db \bar{\phi}(\bar{w}) + i \db \bar{X}(\bar{w}))e^{-\utk \bar{\phi}(\bar{w})} \,,
\ee
where both integrals are taken counter-clockwise.

Writing
expression (\ref{classu}) in Minkowskian cylinder coordinates,  we get the same type
behavior as in the Liouville case. Assuming $p_1<0$ and the boundary conditions
$r(\sigma, t \rightarrow \pm \infty) \rightarrow \infty $,
 the solution behaves, in the far past and future, as
\be
\lim_{t \rightarrow - \infty} u  &=& \lim_{t \rightarrow - \infty} \frac{e^{r + i \theta }}{2 \sqrt{\lambda}} =
e^{\uk (\phi(x^+) + \bar{\phi}(x^-)+ i X(x^+) + i \bar{X}(x^-))} \\
\lim_{t \rightarrow + \infty} u  &=& \lim_{t \rightarrow + \infty} \frac{e^{r + i \theta }}{2\sqrt{\lambda} } =
- \lambda e^{\uk (\phi(x^+) + \bar{\phi}(x^-)+ i X(x^+) + i \bar{X}(x^-))} A(x^+) \bar{A}(x^-) \,.
\ee
We see thus that the incoming and outgoing fields are free fields.
The cigar scatters the incoming free field $\phi(\sigma, t) + i X(\sigma, t)$ into an  outgoing field
which has opposite momentum $-p_1>0$. The case with $p_1<0$ is obtained by time reversal.

\subsection{The reflection coefficient}
The work \cite{Kruger:2004jm} quantized the field $u$
in the Minkowskian cylinder. We will recast here those results
in the Euclidean complex plane\footnote{The main difference between the two
cases lies in the normal ordering of the zero modes and in the fact that
the momentum operator in the complex plane is not Hermitian in the presence
of a background charge (see eq.(\ref{hermitp}) below).
For details on the differences between the quantization
in the cylinder and the complex plane see \cite{Kazama:1993sj}.}.
The quantum free fields have mode expansions
\be
\phi(z) &=& \b{x}_1 -i\frac{\b{p}_1}{2}\log z
+ \frac{i}{\sqrt{2}} \sum_{n \neq 0} \frac{\alpha_n^{(1)}}{n}\frac{1}{z^n} \,,
\label{modphi}
\\
\bar{\phi}(\bar{z}) &=& \b{x}_1 -i\frac{\b{p}_1}{2}\log \bar{z}
+ \frac{i}{\sqrt{2}} \sum_{n \neq 0} \frac{\bar{\alpha}_n^{(1)}}{n} \frac{1}{\bar{z}^n} \,,
\label{modphibar}
\ee
and
\be
X(z) &=& \b{x}_2 -i\frac{\b{p}_2}{2}\log z
+ \frac{i}{\sqrt{2}} \sum_{n \neq 0} \frac{\alpha_n^{(2)}}{n}\frac{1}{z^n} \,,
\\
\bar{X}(\bar{z}) &=& \bb{x}_2 -i\frac{\bb{p}_2}{2}\log \bar{z}
+ \frac{i}{\sqrt{2}} \sum_{n \neq 0} \frac{\bar{\alpha}_n^{(2)}}{n} \frac{1}{\bar{z}^n} \,.
\ee
The zero modes of the $X(z)$ and $\bar{X}(\zb)$ are independent since $X$ is a
compact coordinate, with radius~$2\pi$. The modes
satisfy the commutation relations
\be
[\b{x}_1, \b{p}_1] = i \qquad [ \b{x}_2, \b{p}_2] = i \qquad
  [\bar{\bf x}_2, \bar{\bf p}_2] = i
\ee
\be
[\alpha_n^{(i)}, \alpha_m^{(j)}]= n \, \delta_{n+m,0}\, \delta^{ij} \qquad i,j=1,2
\ee
with all other commutators vanishing.
The field $X$ is compact with radius $\sqrt{k}$, therefore the spectrum of
$\b{p}_2, \bb{p}_2$ is
\be
p_2 &=& \frac{m+nk}{\sqrt{k}} \qquad \bar{p}_2 = \frac{m-nk}{\sqrt{k}} \,.
\label{pedos}
\ee
Normal ordering $::$ is defined as we did in the Liouville case, and the short distance singularities are
\be
\phi(z) \phi(w) &=& :\phi(z) \phi(w): - \frac12 \log(z-w) \,, \\
X(z) X(w)& =& :X(z)X(w): - \frac12 \log(z-w)\,,
\ee
and similarly for $\bar{\phi}(\zb)$ and $\bar{X}(\zb)$.
The stress tensor of the quantum theory is
\be
T(z) = -(\d \phi(z))^2 -b \d^2 \phi(z) - (\d X(z))^2  \,,
\ee
with
\be
b= \frac{1}{\sqrt{k-2}}\,,
\ee
and there is a similar anti-holomorphic copy.
The central charge is
\be
c= 2 + \frac{6}{k-2} \,.
\ee
A chiral vertex operator $:e^{2bj \phi(z)} :$
has holomorphic conformal
dimension
\be
\Delta =- \frac{j(j+1)}{k-2} \,,
\label{cdim}
\ee
and carries $\b{p}_1$ momentum
\be
p_1 = -2ibj \,.
\label{pejota}
\ee
For normalizable states with $j = -\frac12 + i \mathbb{R}$,
the last expression is  consistent
with the anomalous hermiticity  of $\b{p}_1$,
\be
\bp_1^{\dagger} = \bp_1 -2ib\,,
\label{hermitp}
\ee
which follows from the presence of the background charge.

\ni
In order to build the quantum counterpart
of the classical field $u(z,\zb)$ of (\ref{classu}), let us define
\be
\b{f}(z) &=& e^{b \phi(z) + i b\eta X(z)} \,,
\label{efe}
\\
\bb{f}(\zb) &=& e^{b \bar{\phi}(\zb) + i b\eta \bar{X}(\zb)} \,,
\ee
where
\be
\eta = \frac{1}{b \sqrt{k}} = \sqrt{\frac{k-2}{k}}\,,
\ee
and
\be
\d \b{A}(z) &=& \uk ( \d \phi(z) + i \d X(z))e^{-2b \phi(z)} \,, \\
\db \bar{\b{A}}(\zb) &=& \uk ( \db \bar{\phi}(\zb)
+ i \db \bar{X}(\zb))e^{-2b \bar{\phi}(\zb)} \,.
\ee
The change in the exponents of  $\phi, \bar{\phi}$ from the classical case
is in
order for $\d \b{A}, \db \bb{A}$ to be primaries of conformal dimension one.
The monodromy-preserving solutions for $\b{A}$ and $\bb{A}$ are now
\be
\b{A}(z) &=& \frac{1}{\sqrt{k}}\frac{1}{(e^{- 2 b \pi (\b{p}_1-2ib)} -1)} \b{Q}(z) \,,
\label{quantuma} \\
\bar{\b{A}}(\zb) &=& \frac{1}{\sqrt{k}} \bar{\b{Q}}(\zb)
\frac{1}{(e^{- 2 b \pi \b{p}_1} -1)} \,,
\ee
where
\be
\b{Q}(z)&=& \int_z^{ze^{2\pi i}} \!\!\!\!\!\!\! dw \, (\d \phi(w) + i \d X(w))e^{-2b \phi(w)} \,,
\label{kiu}
\\
\bar{\b{Q}}(\zb)&=&
\int_{\zb}^{\zb e^{2\pi i}} \!\!\!\!\!\!\! d \bar{w} \,
(\db \bar{\phi}(\bar{w}) + i \db \bar{X}(\bar{w}))e^{-2b \bar{\phi}(\bar{w})} \,.
\ee
Notice the shift on $\b{p}_1$ in (\ref{quantuma}) when it appears multiplying from the left.
The fields $\b{Q}(z)$ and $\bb{Q}(\zb)$ are primary fields of dimension zero, as in (\ref{dimscreen}).

We can now define the quantum version of $u(z,\zb)$ in (\ref{classu}) as
\be
\b{u}(z,\zb) &=&
\b{f}(z)e^{-b \b{x}_1}\bar{\b{f}}(\zb)
- \lambda \b{A}(z)\b{f}(z) e^{b \b{x}_1} \bar{\b{f}}(\zb)\bar{\b{A}}(\zb) \,,
\nn
\\
&=&
\b{f}(z)e^{-b \b{x}_1}\bar{\b{f}}(\zb)
- \frac{\lambda}{k} \frac{e^{-2\pi i b^2 (\b{j}+1)}}{2i\sin (2\pi b^2(\b{j}+1))}
\b{Q}(z)\b{f}(z) e^{b \b{x}_1} \bar{\b{f}}(\zb)\bar{\b{Q}}(\zb)
\frac{e^{-2\pi i b^2 \b{j}}}{2i\sin (2\pi b^2\b{j})} \,.
\label{opu}
\ee
We have defined the operator $\b{j}$ as (see (\ref{pejota}))
\be
\b{p}_1 = -2ib\b{j} \,,
\ee
and we have added the factor $e^{b \b{x}_1}$ in order to compensate for the
doubling of $\b{x}_1$ in the mode expansion (\ref{modphi})-(\ref{modphibar})
of $\phi$ and $\bar{\phi}$.
The position of  $e^{b \b{x}_1}$ is chosen so that we get the monodromy
\be
\b{u}(ze^{2 \pi i}, \zb e^{-2\pi i}) = \b{u}(z,\zb)e^{\frac{\pi i}{\sqrt{k}}(\b{p}_2-\bb{p}_2)}
\ee
as in the classical solution.

It was shown in \cite{Kruger:2004jm} that the operator $\b{u}$ in the Minkowskian cylinder satisfies
the locality condition
\be
[\b{u}(t,\sigma), \b{u}(t, \sigma')]=0 \,,
\label{loca2}
\ee
similar to (\ref{loca1}).

The above expression for $\b{u}$ involves  the product  of $\b{f}$ and $\b{Q}$ at the same
point. Notice that the first term of the integrand of~$\b{Q}$ in (\ref{kiu})
is a total derivative that can be integrated. Its multiplication
by $\b{f}$ gives zero since
\be
\lim_{z' \rightarrow z}  :e^{-2b\phi(z')}: \b{f}(z)
= \lim_{z' \rightarrow z}   :e^{-2b\phi(z')}\b{f}(z) :(z'-z)^{b^2} = 0\,.
\ee
So for the product $\b{Q}(z)\b{f}(z)$ we only need to compute
\be
\b{Q}(z) \b{f}(z) &=&
\int_z^{ze^{2\pi i}}\!\!\!\!\!\!\!\!\!\!\!\!\!\ dw
\,\, i : \d X(w)e^{-2b \phi(w)}: :e^{b\phi(z) + i b \eta X(z)}: \,, \\
&=& \int_z^{ze^{2\pi i}}\!\!\!\!\!\!\!\!\!\!\!\!\!\ dw
\,\, \left[ i  : \d X(w)e^{-2b \phi(w)}e^{b\phi(z) + i b \eta X(z)}:
(w-z)^{b^2} +
\right.
\nn \\
&& \left. + \,\,
 :e^{-2b \phi(w)}e^{b\phi(z) + i b \eta X(z)}:
\, \frac{\eta}{2b}
\frac{\d}{\d w}(w-z)^{b^2}
\right] \,.
\nn
\ee
The second term  can be integrated by parts and the boundary terms vanish.
The final expression is thus
\be
\b{Q}(z) \b{f}(z) &=&
\int_z^{ze^{2\pi i}}\!\!\!\!\!\!\!\!\!\!\!\!\!\ dw
\,\, :( \eta \d \phi(w) + i \d X(w))e^{-2b \phi(w)} e^{b\phi(z) + i b \eta X(z)}:(w-z)^{b^2} \,,
\ee
and similarly
\be
\bb{f}(\zb)
\bb{Q}(\zb) &=&
\int_{\zb}^{\zb e^{2\pi i}}\!\!\!\!\!\!\!\!\!\!\!\!\!\ d \bar{w}
\,\, :( \eta \db \bar{\phi}(\bar{w}) + i \db \bar{X}(\bar{w}))e^{-2b \bar{\phi}(\bar{w})}
e^{b\bar{\phi}(\zb) + i b \eta \bar{X}(\zb)}:(\zb -\bar{w})^{b^2} \,.
\ee

\vskip 0.5cm
\ni
To compute the reflection coefficient, note that the mapping $j \rightarrow -j-1$ leaves
the conformal dimension (\ref{cdim}) invariant and for normalizable states with $j = -\frac12 + i P$
it amounts to inverting the sign of the momentum $P$.
As in the case of Liouville theory, we introduce
 an operator $\ss{S}$ that maps the state with
momentum $j$ to the state with momentum $-j-1$
\be
\ss{S}|j,m,n\rangle = |-j-1,m,n \rangle\,,
\label{sonj}
\ee
and leaves $\b{u}$ invariant,
\be
\label{uinv}
\ss{S}^{-1}\b{u}(z,\zb)\ss{S} = \b{u}(z,\zb)\,.
\ee
The operator $\ss{S}$ can be decomposed into
\be
\ss{S}= \ss{P}\ss{R}\,,
\ee
where  $\ss{P}$ acts only on the $\b{j}$
momentum of a state and changes its value to~$-j-1$, and $\ss{R}$
acts as
\be
\ss{R}|j,m,n\rangle
= R(j,m,n) |j,m,n\rangle \,,
\ee
with $R(j,m,n)$ being  the reflection coefficient we wish to compute.
Note that from  $\ss{S}^2= 1$ it follows that
\be
R(j,m,n) R(-j-1,m,n)= 1 \,.
\label{invr}
\ee
For the matrix elements of the operator $\b{u}(z,\zb)$ at $z=\zb=1$, we get,
taking proper account of the zero modes in (\ref{opu}),
\be
&& \langle j',m',n' | \b{u}(1,1)|j,m,n \rangle =
\delta_{n',n}\delta_{m',m+1} \left[ \delta(j'-j-\frac12) + \, \delta(j'-j+ \frac12) D(j,m,n) \right] \,.
\label{matrixu}
\ee
We have defined
\be
D(j,m,n)= \frac{\lambda}{4k^2}
( j -\frac12 (m+nk )) (j- \frac12(m-nk) )
\frac{e^{-2 \pi i b^2 (2j+1/2)} I(j) \bar{I}(j)}{\sin(2 \pi b^2 (j+1/2)) \sin(2 \pi b^2 j)}
\ee
and
\be
I(j) &=& \oint dw \, w^{2b^2j-1}(w-1)^{b^2} \,, \\
\bar{I}(j)&=&  e^{-\pi i b^2 }I(j)\,, \nn \\
&=& \oint d \wb \, \wb^{2b^2j-1}(1-\wb)^{b^2} \,.
\ee
Both integrals are taken counterclockwise in the unit circle.
Deforming the contour as we did in the Liouville case, we get
\be
I(j)&=& e^{\pi i b^2}(e^{4\pi i  b^2 j}-1) \int^{1}_0 \!\!dw \, w^{2b^2j-1}(1-w)^{b^2}\,, \nn \\
&=&e^{\pi i b^2}(e^{4\pi i  b^2 j}-1) \frac{\Gamma(2b^2j) \Gamma(b^2 +1)}{\Gamma(b^2(2j+1)+1)},
\ee
where we have used (\ref{euler}).
The final expression for $D(j,m,n)$ is thus
\be
 D(j,m,n)=( j -\frac12 (m+nk )) (j- \frac12(m-nk) )
\frac{ \lambda \Ga^2(b^2+1)}{k^2} \frac{\ga(2 b^2 j)}{\ga(1+b^2(2j+1))} \,.
\ee
From eq.(\ref{uinv}), it follows that the matrix element (\ref{matrixu}) should be equal to
\be
&&  \langle j',m',n' |\ss{S}^{-1}\b{u}(1,1) \ss{S}|j,m,n \rangle =
\delta_{n',n}\delta_{m',m+1}
 \left[ \delta(j'-j+\frac12) R^{-1}(j+ \frac12,m+1,n)R(j,m,n) \, + \right. \nn \\
&& \qquad \left. \delta(j'-j- \frac12)  R^{-1}(j+ \frac12,m+1,n)R(j,m,n) D(-j-1,m,n) \right] \,.
\label{matrixus}
\ee
Comparing (\ref{matrixu}) and (\ref{matrixus}) leads to two functional equations for $R(j,m,n)$,
but using (\ref{invr}) it is easy to see that they are equivalent. The resulting equation is
\be
R(j-\frac12,m+1,n)&=& R(j,m,n) D^{-1}(j,m,n)\,.
\label{firstpair2}
\ee
This equation is not enough to fix  $R(j,m,n)$, since
any solution can be multiplied by a periodic function of $j$, with period $\frac12$, and/or a periodic function
of $m$ with period $1$.

\subsection{The sine-Liouville dual}
In light of the FZZ duality, it is natural
to expect that the reflection coefficient
$R(j,m,n)$ will get fixed through a second set of equations, associated to a scattering problem
of sine-Liouville type.
On the other hand, we do not expect that
such a scattering problem corresponds to   solving  the equations of motion of the classical
sine-Liouville Lagrangian. The reason is that the sine-Liouville
interaction carries one unit of winding, and the reflection coefficient,
which is equal to the two-point function, conserves the winding number \cite{FZZ}.

To arrive to  the form of the sine-Liouville scattering that we need, note
that we expect  the FZZ duality to be  related to a $b \leftrightarrow b^{-1}$
transformation. Even though, unlike Liouville, the theory is not self-dual,  the exact solution
of the closely related $H_3$ WZW model shows that the two functional equations that determine the three-point functions
are related by a $b \leftrightarrow b^{-1}$ transformation~\cite{Teschner:1997ft}. Applying this transformation
to $\b{f}(z)$ in (\ref{efe}), we get
\be
\bg(z) = e^{\frac{1}{b} \phi(z)+ i\frac{\eta}{b} X(z)} \,.
\ee
In order to invert the sign of the $\b{p}_1$ momentum carried by $\bg$ via scattering,
it follows from (\ref{lagrangian}) that we need {\it two}
sine-Liouville interactions. This in turn allows to conserve the winding number by adjoining two
sine-Liouville interactions with opposite winding.
We propose that the FZZ dual of $\b{u}(z,\zb)$ is the field
\be
\bv(z,\zb) &=& \bg(z)e^{-\frac{1}{b}\b{x}_1} \bar{\bg} (\zb)  \\
&& \qquad + \frac{\tilde{\lambda}}{(e^{4 \pi i \b{j}}-1)(e^{2 \pi i (\b{j} - \b{r} +k' )}-1)}
\b{B}(z)\bg(z)e^{\frac{1}{b}\b{x}_1} \bar{\bg}(\zb) \bar{\b{B}}(\zb)
\frac{1}{(e^{4 \pi i \b{j}}-1)(e^{2 \pi i (\b{j} - \bar{\b{r}} -k')}-1)} \,,
\nn
\ee
where $\b{r}= \frac{\sqrt{k}}{2}\b{p}_2, \bar{\b{r}}= \frac{\sqrt{k}}{2}\bar{\b{p}}_2  $ and
\be
\b{B}(z)  = \int_{z}^{ze^{2\pi i}} \!\!\!\!\!\!\!\!\!\!\!\!\!\ dw_1 \,\,  e^{i\sqrt{k} X(w_1) - \frac{1}{b} \phi(w_1)}
\int_{z}^{ze^{2\pi i}} \!\!\!\!\!\!\!\!\!\!\!\!\!\ dw_2  \,\, e^{-i\sqrt{k} X(w_2) - \frac{1}{b} \phi(w_2)}\,,
\ee
and similarly for $\bar{\b{B}}(\zb)$.
The main evidence for our proposal is that this expression yields the
correct reflection coefficient and a special structure constant which we discuss in the next section.
Much as in the Liouville or the cigar cases, the classical form of $\bv(z,\zb)$ maps an incoming
free field into an outgoing free field when expressed in Minkowskian coordinates in the cylinder.

The product of $\b{B}(z)$ and $\bg(z)$ is given by
\be
\b{B}(z) \bg(z) = \int_{z}^{ze^{2\pi i}} \!\!\!\!\!\!\!\!\!\!\!\!\!\ dw_1
\int_{z}^{ze^{2\pi i}} \!\!\!\!\!\!\!\!\!\!\!\!\!\ dw_2
&& :e^{ - \frac{1}{b} \phi(w_1) - \frac{1}{b} \phi(w_2) + \frac{1}{b} \phi(z)}
e^{i\sqrt{k} X(w_1) - i\sqrt{k} X(w_2) + i\frac{\eta}{b} X(z)}:  \nn \\
&& \qquad \qquad \times (w_1-z)^{k'} (w_1-w_2)^{-k+1}
\ee
and similarly for $\bar{\bg}(\zb) \bar{\b{B}}(\zb)$.
The spectrum of $\b{r}, \bar{\b{r}}$ is  (see (\ref{pedos}))
\be
r &=& \frac{m + nk}{2} \,,\\
\bar{r} &=& \frac{m-nk}{2} \,.
\ee
We can now consider a generic matrix element of $\bv(1,1)$, which gives
\be
\langle j',m',n'| \bv(1,1) |j,m,n \rangle = \delta_{n',n} \delta_{m',m+k'}
\left[ \delta(j'-j-\frac{k'}{2}) + E(j,m,n) \delta(j'-j+\frac{k'}{2})   \right]
\label{matrixlv}
\ee
where
\be
E(j,m,n) =  \frac{ \tilde{\lambda} K(j,r) \bar{K}(j,\bar{r})}{(e^{2 \pi i (2j-k')} - 1 )
(e^{4 \pi i j}-1)(e^{2 \pi i (j-r)}-1)(e^{2 \pi i (j-\bar{r}-k')}-1)} \,.
\ee
Here $\bar{K}(j,\bar{r}) = e^{-i\pi k'} K(j,r) \big|_{r=\bar{r}}$ and
\be
K(j,r) &=& \oint \oint dw_1  dw_2  (w_1-1)^{k'} (w_1-w_2)^{-k+1} w_1^{j+r} w_2^{j-r}  \\
&=& \oint  dw_1    w_1^{2j-k'} (w_1-1)^{k'}    \oint dy  y ^{j-r} (1-y)^{-k+1} \nn \\
&=& e^{i \pi  k'} (e^{2 \pi i(2j-k')}-1)
(e^{2 \pi i(j-r)}-1)
 \int_0^1  \! dw_1 w_1^{2j-k'} (w_1-1)^{k'}
\int_0^1 \!  dy  y ^{j-r} (1-y)^{-k+1} \nn \\
&=&  e^{i \pi  k'} (e^{2 \pi i(2j-k')}-1)
(e^{2 \pi i(j-r)}-1) \frac{\Ga(2j-k'+1)\Ga(k'+1)}{\Ga(2j+2)}
\frac{\Ga(j-r+1)\Ga(-k')}{\Ga(j-r-k'+1)} \,, \nn
\ee
where in the first two lines the integrals are taken counterclockwise in the unit circle,
and in the second line we changed variables to $(y=w_2/w_1,w_1)$.
The contour deformations in going from the second to the third line
are the same as  in the previous cases.
The final expression for $E(j,m,n)$ is
\be
E(j,m,n) = \frac{\tilde{\lambda}\pi^2}{\sin^2(\pi k')} \ga(2j-k'+1) \ga(-2j-1) \frac{\Ga(j-r + 1)}{\Ga(j-r-k'+1)}
\frac{\Ga(-j+ \bar{r} + k')}{\Ga(-j+\bar{r})}\,.
\ee
As before, to obtain the sought-for equations for $R(j,m,n)$ we impose the condition
\be
\langle j',m',n'| \bv(1,1) |j,m,n \rangle = \langle j',m',n'| \ss{S}^{-1} \bv(1,1) \ss{S} |j,m,n \rangle
\ee
and this yields two equations for $R(j,m,n)$, which are equivalent after using (\ref{invr}).
The resulting equation is
\be
R(j-\frac{k'}{2},m+k',n)&=& R(j,m,n) E^{-1}(j,m,n)\,.
\label{seconddiff}
\ee
This equation, along with (\ref{invr})  and (\ref{firstpair2}),
has  as a  solution
\be
R(j,m,n) = \nu^{2j+1}
\frac{\Ga(-2j-1)}{\Ga(2j+1)}
\frac{\Ga(1-b^2(2j+1))}{\Ga(1+b^2(2j+1))}
\frac{\Ga(j+1-\frac12(m+nk))}{\Ga(-j-\frac12(m+nk))}
\frac{\Ga(j+1+\frac12(m-nk))}{\Ga(-j+\frac12(m-nk))}
\nn
\\
\,\,
\ee
where $\nu=-\lambda \Ga^2(b^2)/k^2 = (\tilde{\lambda}\pi^2 /k'^2 \sin^2(\pi k'))^{b^2}$. This result
coincides with the reflection coefficient for the cigar obtained with other methods (see e.g. \cite{Mukherjee:2006zv}).

\subsection{Relation with parafermionic symmetry}
The two-dimensional black hole has two holomorphic and two
anti-holomorphic parafermionic conserved currents,
which come from the affine symmetries of the $SL(2,R)$ WZW model.
In terms of the fields~$\phi, X$, the currents are
\be
\label{parafermions}
\psi^{\pm}= (i\sqrt{k}\partial X \mp \sqrt{k-2}\partial \phi )
e^{\mp \frac{2i}{\sqrt{k}}X} \,.
\ee
In the work \cite{Mukherjee:2006zv}, following \cite{Teschner:1995yf,Giveon:2001up},
the reflection coefficient $R(j,m,n)$ was obtained using properties of degenerate operators
of the parafermionic symmetry.
Remarkable, the method of \cite{Mukherjee:2006zv}  to determine the reflection coefficient,
leads to the same two difference equations (\ref{firstpair2}) and (\ref{seconddiff}).

In this section we would
like to point out some connections between the approach in \cite{Mukherjee:2006zv} and the methods used in this paper.

Firstly, note that the screening charges $\b{Q}(z)$ and $\b{B}(z)$ which appear in $\b{u}$ and $\b{v}$ are built from primary operators of dimension one,
which, as shown in \cite{Mukherjee:2006zv}, commute with the parafermionic generators.

The primary fields of the parafermionic symmetry can be written as
$V_{j,r,\bar{r}}$.
The method of \cite{Mukherjee:2006zv} exploits the fact that the operators $V_{\frac12,\frac12,\frac12}$
and $V_{\frac{k'}{2},\frac{k'}{2},\frac{k'}{2}}$
are degenerate operators of the parafermionic algebra.\footnote{The conventions in \cite{Mukherjee:2006zv} differ
from those of this paper by the replacements $j,m,\bar{m} \rightarrow -j,r,\br$.}

One can check from the  free field representation of the $SL(2,R)/U(1)$ primaries given in~\cite{Mukherjee:2006zv},
that the free field representations of these two
degenerate fields coincide with the first terms in  $\b{u}(z,\bar{z})$ and $\b{v}(z,\bar{z})$,
namely their asymptotic value  in the far past.
The free field representation
of the primaries is valid when the interaction in the $SL(2,R)/U(1)$ Lagrangian is turned off.
This  
 suggests the identifications
\be
&& \b{u}(z,\bar{z}) = V_{\frac12,\frac12,\frac12} \,,
\label{id1}
\\
&& \b{v}(z,\bar{z}) = V_{\frac{k'}{2},\frac{k'}{2},\frac{k'}{2}} \,,
\label{id2}
\ee
in the fully interacting theory. To prove  these identities, note that
in the OPE of  these two degenerate fields with a generic primary of $SL(2,R)/U(1)$,
we have  the fusion rules \cite{Awata:1992sm}
\begin{eqnarray}
V_{\frac12,\frac12,\frac12} V_{j,r,\br} & \sim &
C^+_{j,r,\br} \left[ V_{j+\frac12,r+\frac12,\br+\frac12} \right]
+ C^{-}_{j,r,\br}\left[V_{j-\frac12, r+\frac12,\br+\frac12}\right]\,,
\label{fusion}
\ee
and
\be
V_{\frac{k'}{2},\frac{k'}{2},\frac{k'}{2}} V_{j,r,\br} \sim
\tilde{C}^+_{j,r,\br} \left[ V_{j+\frac{k'}{2},r
+\frac{k'}{2},\br+\frac{k'}{2}} \right]
+ \tilde{C}^{-}_{j,r,\br}\left[V_{j-\frac{k'}{2},
r+\frac{k'}{2},\br+\frac{k'}{2}}\right]
\nn
+ \, \tilde{C}^{\times}_{j,r,\br}\left[V_{-\frac{k'}{2}-j-1,
r+\frac{k'}{2},\br+\frac{k'}{2}}\right] \,\,.
\\
\label{fusion2}
\end{eqnarray}
The fields can be normalized so that $C^+_{j,r,\br}= \tilde{C}^+_{j,r,\br} =1$.
The first remarkable connection between the approach of this paper
and that  of \cite{Mukherjee:2006zv},
are the identities
\be
C^{-}_{j,r,\br} &=& D(j,m,n)\,,
\label{cd}
\\
\tilde{C}^{-}_{j,r,\br} &=& E(j,m,n) \,,
\label{ce}
\ee
which will help us to establish (\ref{id1})-(\ref{id2}).

Consider the field $V_{\frac12,\frac12,\frac12}$ first. Taking the
$\langle j', m', n'|\cdot |0\rangle$
matrix element of both
sides of eq.(\ref{fusion}) evaluated at
$V_{\frac12,\frac12,\frac12}(1) V_{j,r,\br}(0)$,
we get on the r.h.s, using (\ref{cd}), precisely eq.(\ref{matrixu}).
Therefore, the generic matrix elements of $\b{u}$ and $V_{\frac12,\frac12,\frac12}$
coincide, and this establishes their identity as operators.

For the field $V_{\frac{k'}{2},\frac{k'}{2},\frac{k'}{2}}$,
we should  notice that the first and last term in the r.h.s of (\ref{fusion2})
should be treated as the incoming and reflected wave function of the same field,
since they are related by $j \rightarrow -j -1$.
We can define
\be
 \tilde{V}_{j+\frac{k'}{2},r +\frac{k'}{2},\br+\frac{k'}{2}} \equiv
V_{j+\frac{k'}{2},r +\frac{k'}{2},\br+\frac{k'}{2}}
 + R^{-1}(j+{k'}/{2}, m+ k',n)  V_{-\frac{k'}{2}-j-1,r+\frac{k'}{2},\br+\frac{k'}{2}} \,,
\label{vtilde}
\ee
where we have identified
\be
R^{-1}(j+{k'}/{2}, m+ k',n) =  \tilde{C}^{\times}_{j,r,\br} \,,
\ee
and we should identify the state $|j+\frac{k'}{2},m+k',n \rangle$ with the action of
$\tilde{V}_{j+\frac{k'}{2},r+\frac{k'}{2},\br+\frac{k'}{2}}$ on the vacuum~$|0 \rangle$.
Note that this is consistent with eqs.(\ref{sonj})-(\ref{invr}).
Now eq.(\ref{fusion2}) can be rewritten as
\be
V_{\frac{k'}{2},\frac{k'}{2},\frac{k'}{2}} V_{j,r,\br} &\sim&
\left[ \tilde{V}_{j+\frac{k'}{2},r+\frac{k'}{2},\br+\frac{k'}{2}} \right]
+ \tilde{C}^{-}_{j,r,\br}\left[V_{j-\frac{k'}{2}, r+\frac{k'}{2},\br+\frac{k'}{2}}\right] \,.
\ee
Taking the $\langle j', m', n'|\cdot |0\rangle$ matrix element on both sides of this  equation,
and using (\ref{matrixlv}) and~(\ref{ce}), it follows that the generic
matrix elements of $\b{v}$ and $V_{\frac{k'}{2},\frac{k'}{2},\frac{k'}{2}}$ coincide, and this proves the identity~(\ref{id2}).

\vskip 0.5 cm
In order to fully establish the validity of the field
$\b{v}$ as the dual to  $\b{u}$, it must be shown that in
the Minkowskian cylinder a locality property for $\b{v}$ similar to (\ref{loca1}) and (\ref{loca2}) holds,
as well as locality between $\b{v}$ and $\b{u}$.
But this follows from eqs.(\ref{id1}) and (\ref{id2}), and the fact that the locality properties are valid for
the~$V_{\frac12, \frac12, \frac12}$ and~$V_{\frac{k'}{2},\frac{k'}{2},\frac{k'}{2}}$ operators in the $H_3^+$ WZW model,
and survive in the parafermionic theory obtained as the coset~$H_3^+/U(1)$.

\section*{Acknowledgements}
We thank Brenno Carlini-Vallilo, Vladimir Korepin, William Linch and Sunil Mukhi
for conversations. Special thanks to the JHEP referee for many  insightful suggestions which
were essential for the results in Section 3.4.
This work is supported by the Simons Foundation.
\appendix

\section{Useful formulae}
\be
\Ga(x)\Ga(1-x) &=& \frac{\pi}{\sin(\pi x)} \\
\ga(x) &=& \frac{\Gamma(x)}{\Gamma(1-x)}
\end{eqnarray}
\be
\int_0^1 \!\!\!\! dw \, w^{\alpha-1}(1-w)^{\beta -1} =
\frac{\Gamma(\alpha) \Gamma(\beta)}{\Gamma(\alpha+\beta)}
\label{euler}
\ee


\bibliography{fzzop}

\bibliographystyle{JHEP}

\end{document}